\documentclass[conference]{IEEEtran}
\IEEEoverridecommandlockouts
\usepackage{cite}

\usepackage{times}
\usepackage{fancyhdr}
\usepackage[linesnumbered,ruled]{algorithm2e}

\usepackage{amsmath,amssymb,amsfonts}
\usepackage{algorithmic}
\usepackage{placeins}
\usepackage{graphicx}
\usepackage{subfigure}
\usepackage{textcomp}
\usepackage{xcolor}
\usepackage{lscape}

\def\BibTeX{{\rm B\kern-.05em{\sc i\kern-.025em b}\kern-.08em
    T\kern-.1667em\lower.7ex\hbox{E}\kern-.125emX}}
\begin{document}

\title{RegTraffic: A Regression Based Traffic Simulator for Spatiotemporal Traffic Modeling, Simulation and Visualization\\



}

\author{\IEEEauthorblockN{Sifatul Mostafi, Taghreed Alghamdi, Khalid Elgazzar}
\IEEEauthorblockA{\textit{IoT Research Lab, ECSE, Ontario Tech University, Oshawa, ON, Canada}\\
\{sifatul.mostafi, Taghreed Alghamdi, khalid.elgazzar\}@ontariotechu.ca}
}

\maketitle

\begin{abstract}

Traffic simulation is a great tool to demonstrate complex traffic structures which can be extremely useful for the planning, development, and management of road traffic networks. Current traffic simulators offer limited features when it comes to interactive and adaptive traffic modeling. This paper presents RegTraffic, a novel interactive traffic simulator that integrates dynamic regression-based spatiotemporal traffic analysis to predict congestion of intercorrelated road segments. The simulator models traffic congestion of road segments depending on neighboring road links and temporal features of the dynamic traffic flow. The simulator provides a user-friendly web interface to select road segments of interest, receive user-defined traffic parameters, and visualize the traffic for the flow of correlated road links based on the user inputs and the underlying correlation of these road links. Performance evaluation shows that RegTraffic can effectively predict traffic congestion with a Mean Squared Error of 1.3 Km/h and a Root Mean Squared Error of 1.71 Km/h. RegTraffic can effectively simulate the results and provide visualization on interactive geographical maps.

\end{abstract}

\begin{IEEEkeywords}
Road traffic, simulator, regression, visualization, software
\end{IEEEkeywords}

\section{Introduction}\label{sec:Introduction}

\par With the advancement of computer technologies and software engineering, computer-based traffic simulation has become a popular approach for traffic analysis in support of the evaluation and design of Intelligent Transport Systems (ITS) \cite{Barcelo2010}. Traffic simulation software supported by the ability to emulate the variability of spatial and temporal components in traffic flows is a practical tool for capturing and explaining complex traffic systems \cite{Ejercito2017}.

The purpose of developing Traffic simulation tools is to experiment with varieties of strategies in traffic modeling \cite{Pell2017}. Traffic simulation software tools and models built on real-life traffic data are widely applied to support real-time traffic decisions and management solutions.

\par 
Regression analysis in the traffic domain is a well-established approach that facilitates traffic modeling and prediction \cite{Yan2009}. Regression-based traffic modeling helps in analyzing complex traffic structures which is a useful method for the development and planning of traffic systems and networks.  Hence, traffic congestion estimation and computerized simulation are suitable options for policymakers to analyze different complex traffic scenarios and take actions accordingly \cite{Mubasher2015}.

\par A lot of microscopic and macroscopic traffic simulators have been developed including SUMO \cite{Lopez2010}, Aimsun \cite{Casas2010}, Traffsim \cite{Lindorfer2017}, SUMMIT \cite{Cai2020}, SifTraffic \cite{Sorenson2021} and VISSIM \cite{Fellendorf2010}. These simulators have practical use cases in traffic analysis including traffic flow measurement, multi-agent simulation, particle-based simulation, and so on. Although these state-of-the-art simulators have many practical traffic use cases, they face challenges in the application of simulating road traffic congestion in heterogeneous road transportation networks with a small amount of real-time data \cite{Pell2017}. Also, these simulators lack the feature to adopt regression-based traffic modeling and simulate traffic congestion of a road link depending on traffic congestion of neighboring road links. Some of the simulators do not provide visualization for the simulation results in interactive geographical maps.

\par We aim to develop a regression-based traffic simulator for spatiotemporal traffic modeling to predict traffic congestion of a road link depending on neighboring road segments and provide features to simulate and visualize the results using interactive geographical maps.
 
\par The remainder of this paper is organized as follows. Section \ref{sec:BackgroundAndRelatedWork} briefly reviews the state-of-the-art traffic simulators and their scope in traffic modeling and simulation. The traffic modeling approach of RegTraffic is described in Section \ref{sec:Modeling}. Section \ref{sec:ProcessingPipeline} outlines the processing pipeline of the different components of RegTraffic. Section \ref{sec:Simulation} provides a step-by-step simulation scenario of a traffic use case. Section \ref{sec:PerformanceEvaluation} shows the performance analysis of RegTraffic. Lastly, Section \ref{sec:Conclusion} concludes this work and provides future research directions.

\section{Background and Related Work}\label{sec:BackgroundAndRelatedWork}

\begin{table*}[t]
\def\arraystretch{1.75}%
\caption{Comparison of Traffic Simulators}
\label{table:Comparison of Traffic Simulators}
  \centering
  \resizebox{\textwidth}{!}{
  \begin{tabular}{*{9}{c|c|c|c|c|c|c|c|c}}
    \hline
    Comparison Category Simulator & RegTraffic & FreeSim \cite{Miller2007} & SUMO \cite{Lopez2010} & Aimsun \cite{Casas2010} & TraffSim \cite{Lindorfer2017} & SUMMIT \cite{Cai2020} & SimTraffic \cite{Sorenson2021} & VISSIM \cite{Fellendorf2010} \\
    \hline
    Spatiotemporal Traffic Modeling & Yes & Yes & Yes & Yes & Yes & Yes & Yes & Yes \\
    \hline
    Regression Modeling & Yes & No & No & No & No & No & No & No \\
    \hline
    \shortstack{Interactive Geographical Maps} & Yes & No & Yes & Yes & No & Yes & Yes & Yes \\
    \hline
    Web Interface & Yes & No & No & No & No & No & No & No \\
    \hline
  \end{tabular}
  }
\end{table*}

\par Traffic simulator software is commonly divided into two categories: microscopic traffic simulators \cite{Fellendorf2010,Lopez2010} and macroscopic traffic simulators \cite{Ramadhan2017}. 

\par FreeSim \cite{Miller2007} is a traffic simulator designed to conduct real-time freeway traffic simulation. SUMO (Simulation of Urban Mobility) \cite{Lopez2010} is a microscopic traffic simulator that is developed to process complex and large road networks. SUMO is widely used in many applications including traffic flow modeling \cite{Haddouch2017} and color mapping Google Maps routes \cite{Mena2019}. Aimsun \cite{Casas2010} is a traffic simulator for modeling smart mobility. Traffsim \cite{Lindorfer2017} simulator is widely used for modeling isolated traffic control strategies. SUMMIT \cite{Cai2020} provides functionalities to simulate urban driving in large traffic scenarios with massive and mixed traffic. SifTraffic \cite{Sorenson2021} is a practical software tool to conduct simulations of practical traffic applications. VISSIM \cite{Fellendorf2010} is a microscopic traffic simulator for behavior-based multi-purpose traffic flow simulation.

\par Wang et al. \cite{Wang2017} explored different methods of correcting the traffic simulation models based on linear regression. Golovnin et al. \cite{Golovnin2019} took a web-oriented approach to simulate road traffic, especially in urban settings. Mizuta et al. \cite{Mizuta2015} evaluated the traffic flow near intersections of a metropolitan city to understand how agent-based traffic simulators work to approximate vehicle behaviors.

\par A comparison among the existing traffic simulators along with RegTraffic is listed in Table \ref{table:Comparison of Traffic Simulators} in terms of some key characteristics and features.

\section{Mathematical Modeling}\label{sec:Modeling}

\subsection{Spatial Feature}\label{subsec:ModelingSpatialFeature}

\begin{figure}[ht]
    \centering
    \includegraphics[width=0.8\linewidth]{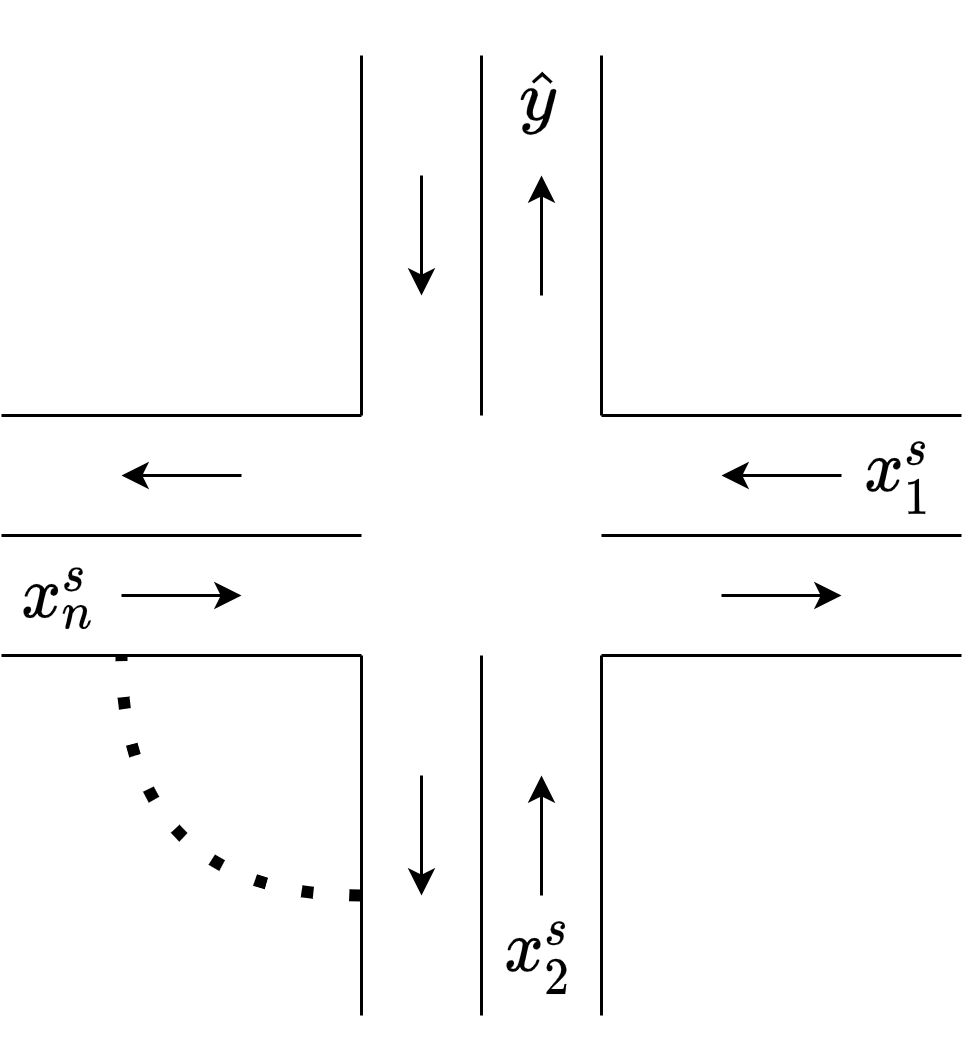}
    \caption{Traffic Road intersection}
    \label{fig:Chapter5TrafficRoadJunction}
\end{figure}

Figure \ref{fig:Chapter5TrafficRoadJunction} shows a traffic road intersection. In this intersection, we consider a road link as the spatial road feature that is dependent on one or several connected spatial road features. For example, for the intersection shown in Figure \ref{fig:Chapter5TrafficRoadJunction}, the road link $\hat{y}$ is a spatial feature and is modeled as a dependent variable in our regression modeling. The road links $x_{1}^{s}$, $x_{2}^{s}$ up to the road link $x_{n}^{s}$ are  the independent spatial features. It’s worth noting that the dependent road link $\hat{y}$ is an outbound while all the independent road links $x_{1}^{s}, x_{2}^{s},...,x_{n}^{s}$ inbound to the intersection. Our proposed traffic modeling approach described in \cite{Sifatul2021} indicates that the dependent spatial feature must be an outbound road link and the independent spatial features must be inbound road links. The model incorporates a set of temporal features that can be extracted from both independent and dependent spatial features through exploratory data analysis. The specific number of temporal features and independent spatial features are arbitrary and dependent on the specific road intersection and their orientation.

Here, $X_{S}$ is defined as a set of independent spatial features $\left \{ x_{1}^{s}, x_{2}^{s},.., x_{n_{s}}^{s} \right \}$ as shown in Eq. \eqref{eqn:XS}.

\begin{equation} \label{eqn:XS}
\begin{aligned}
X_{S} = \left \{ x_{1}^{s}, x_{2}^{s},.., x_{n_{s}}^{s} \right \}
\end{aligned}
\end{equation}

The cardinality of set $X_{S}$ is defined as $n_{s}$ as shown in Eq. \eqref{eqn:NS}. 

\begin{equation} \label{eqn:NS}
\begin{aligned}
n_{s} = \left | X_{S} \right |
\end{aligned}
\end{equation}

\subsection{Temporal Feature Extraction}\label{subsec:ModelingTemporalFeatureExtraction}

In our modeling, we convert temporal features into categorical features using one hot encoding. To simplify our modeling, temporal features are encoded using only two values. Here, $X_{T}$ is a set of temporal features $\left \{ x_{1}^{t}, x_{2}^{t},.., x_{n_{t}}^{t} \right \}$ as shown in Eq. \eqref{eqn:XT}.

\begin{equation} \label{eqn:XT}
\begin{aligned}
X_{T} = \left \{ x_{1}^{t}, x_{2}^{t},.., x_{n_{t}}^{t} \right \}
\end{aligned}
\end{equation}

The cardinality of set $X_{T}$ is defined as $n_{t}$ as shown in Eq. \eqref{eqn:NT}.

\begin{equation} \label{eqn:NT}
\begin{aligned}
n_{t} = \left | X_{T} \right |
\end{aligned}
\end{equation}

The set of temporal features is extracted from spatial features using exploratory data analysis. Here, $X_{T}$ is the output of function $f$ which takes in the set of spatial features $X_{S}$ as input. The function $f$ is a many to many function that takes in a set of spatial features and conducts exploratory data analysis to extract a set of temporal features as shown in Eq. \eqref{eqn:F}

\begin{equation} \label{eqn:F}
\begin{aligned}
X_{T} = f_{ n_{s} \rightarrow n_{t} } (X_{S})
\end{aligned}
\end{equation}

We define the set $X$ as a union of the temporal features $X_{T}$ and spatial features $X_{S}$ as shown in Eq. \eqref{eqn:X}.

\begin{equation} \label{eqn:X}
\begin{aligned}
X = X_{T} \cup X_{S} 
\end{aligned}
\end{equation}

\subsection{Regression Modeling}\label{subsec:RegressionModeling}

\subsubsection{Regression Formation}\label{subsubsec:RegressionFormation}

RegTraffic forms a regression model through a linear combination of both temporal and spatial explanatory features to explain the dependent spatial feature $\hat{y}$ as shown in Eq. \eqref{eqn:Y}. In this equation, all the independent features are associated with their corresponding regression coefficient. $\alpha$ indicates the bias and $\epsilon$ refers to the error term.

\begin{equation} \label{eqn:Y}
\begin{aligned}
\hat{y} = \sum_{i=1}^{n_{t}} \beta_{i}^{t} x_{i}^{t} + \sum_{i=1}^{n_{s}} \beta_{i}^{s} x_{i}^{s} + \alpha + \epsilon 
\end{aligned}
\end{equation}

In the regression Eq. \eqref{eqn:Y}, every explanatory temporal feature from setting $X_{T}$ is associated with a regression coefficient from set $\beta_{T}$ as shown in Eq. \eqref{eqn:BT}.

\begin{equation} \label{eqn:BT}
\begin{aligned}
\beta_{T} = \left \{ \beta_{1}^{t}, \beta_{2}^{t},.., \beta_{n_{t}}^{t} \right \}
\end{aligned}
\end{equation}

Similarly, in the regression Eq. \eqref{eqn:Y}, every explanatory spatial feature from set $X_{S}$ is associated with a regression coefficient from set $\beta_{S}$ as shown in Eq. \eqref{eqn:BS}.

\begin{equation} \label{eqn:BS}
\begin{aligned}
\beta_{S} = \left \{ \beta_{1}^{s}, \beta_{2}^{s},.., \beta_{n_{s}}^{s} \right \}
\end{aligned}
\end{equation}

Here, $\beta$ is defined as the union of set $\beta_{T}$ and $\beta_{S}$

\begin{equation} \label{eqn:B}
\begin{aligned}
\beta = \beta_{T} \cup \beta_{S} 
\end{aligned}
\end{equation}

\subsubsection{Posterior Probability Distribution}\label{subsubsec:PosteriorProbabilityDistribution}

\par We use a novel Bayesian linear regression approach for spatiotemporal traffic modeling of a road link proposed in \cite{Sifatul2021}. Bayesian linear regression formulates a posterior probability distribution of the model parameters rather than just finding a single point estimate. The response variable is drawn from a probability distribution instead of a single value estimation. A Bayesian linear regression model samples the response variable from a normal distribution as shown in Eq. \eqref{eqn:BayesianRegression}.

\begin{equation} \label{eqn:BayesianRegression}
y \sim N(\beta^TX, \sigma^2I)
\end{equation}

\par In Eq. \eqref{eqn:BayesianRegression}, the response variable $y$ is generated from a Gaussian normal distribution, which is characterized by a mean and variance. Eq. \eqref{eqn:Chapter5BayesianProbability} refers to the Bayes Theorem which is the fundamental building block of Bayesian linear regression. Here, $P(\beta\mid \hat{y},X)$ is the posterior probability distribution of the model parameters, $P(\hat{y}\mid\beta,X)$ is the likelihood of the data, $P(\beta \mid X)$ is the prior probability of the parameters and $P(\hat{y}\mid X)$ is the normalization constant. The posterior distribution of the model parameters is proportional to the multiplication of the likelihood of the data and the prior probability of the parameters. A detailed description of the model is described in  \cite{Sifatul2021}.

\begin{equation} \label{eqn:Chapter5BayesianProbability}
P(\beta\mid \hat{y},X) = \frac{P(\hat{y}\mid\beta,X) * P(\beta \mid X)}{P(\hat{y}\mid X)}
\end{equation}

Once the regression model is built, the user can provide new observations for independent spatial features $X_{S}$ and independent temporal features $X_{T}$ into the model. Based on the new observation, the model incorporates the regression coefficients associated with the explanatory variables and predicts the output for the dependent variable $\hat{y}$. An event can be associated with a specific value as an input for any independent spatial feature.

\subsection{Event Integration}\label{subsec:Event Integration}

Here, $X_{E}$ is defined as a set of events $\left \{ X_{1}^{E}, X_{2}^{E},.., X_{n_{E}}^{E} \right \}$ as shown in Eq. \eqref{eqn:XE}.

\begin{equation} \label{eqn:XE}
\begin{aligned}
X_{E} = \left \{ X_{1}^{E}, X_{2}^{E},.., X_{n_{E}}^{E} \right \}
\end{aligned}
\end{equation}

The cardinality of set $X_{E}$ is defined as $n_{E}$ as shown in Eq. \eqref{eqn:NE}. 

\begin{equation} \label{eqn:NE}
\begin{aligned}
n_{E} = \left | X_{E} \right |
\end{aligned}
\end{equation}

After event integration, the independent spatial features associated with an event are integrated into Eq. \eqref{eqn:Y}. If any independent spatial feature is associated with an event, we need to replace the value for the independent spatial feature $x^{S}$ with the events $x^{E}$. Spatial features which are not affected by any specific event are represented by $x^{S\prime}$ along with their model parameter $\beta^{S\prime}$ as shown in Eq. \eqref{eqn:YE}. The amount of spatial features unaffected by any specific event is denoted as shown in Eq. \eqref{eqn:SMiNUSE}.

\begin{equation} \label{eqn:SMiNUSE}
\begin{aligned}
n_{S\prime} = n_{S} - n_{E}
\end{aligned}
\end{equation}

\begin{equation} \label{eqn:YE}
\begin{aligned}
\hat{y} = \sum_{i=1}^{n_{t}} \beta_{i}^{t}
x_{i}^{s} + \sum_{i=1}^{n_{S\prime}} \beta_{i}^{S\prime} x_{i}^{S\prime} + \sum_{i=1}^{n_{E}} \beta_{i}^{E} x_{i}^{E} + \alpha + \epsilon
\end{aligned}
\end{equation}

However, we add a time constraint in association with the temporal components for adding a specific event into the regression equation. For any specific event $X_{E}$ occurred at time $T_E$, the value of spatial feature $X_{S}$ will be replaced by the value of $X_{E}$ if $T_E \subset X_{T}$.

\section{Processing Pipeline}\label{sec:ProcessingPipeline}

\begin{figure}[t]
    \includegraphics[width=1.0\linewidth]{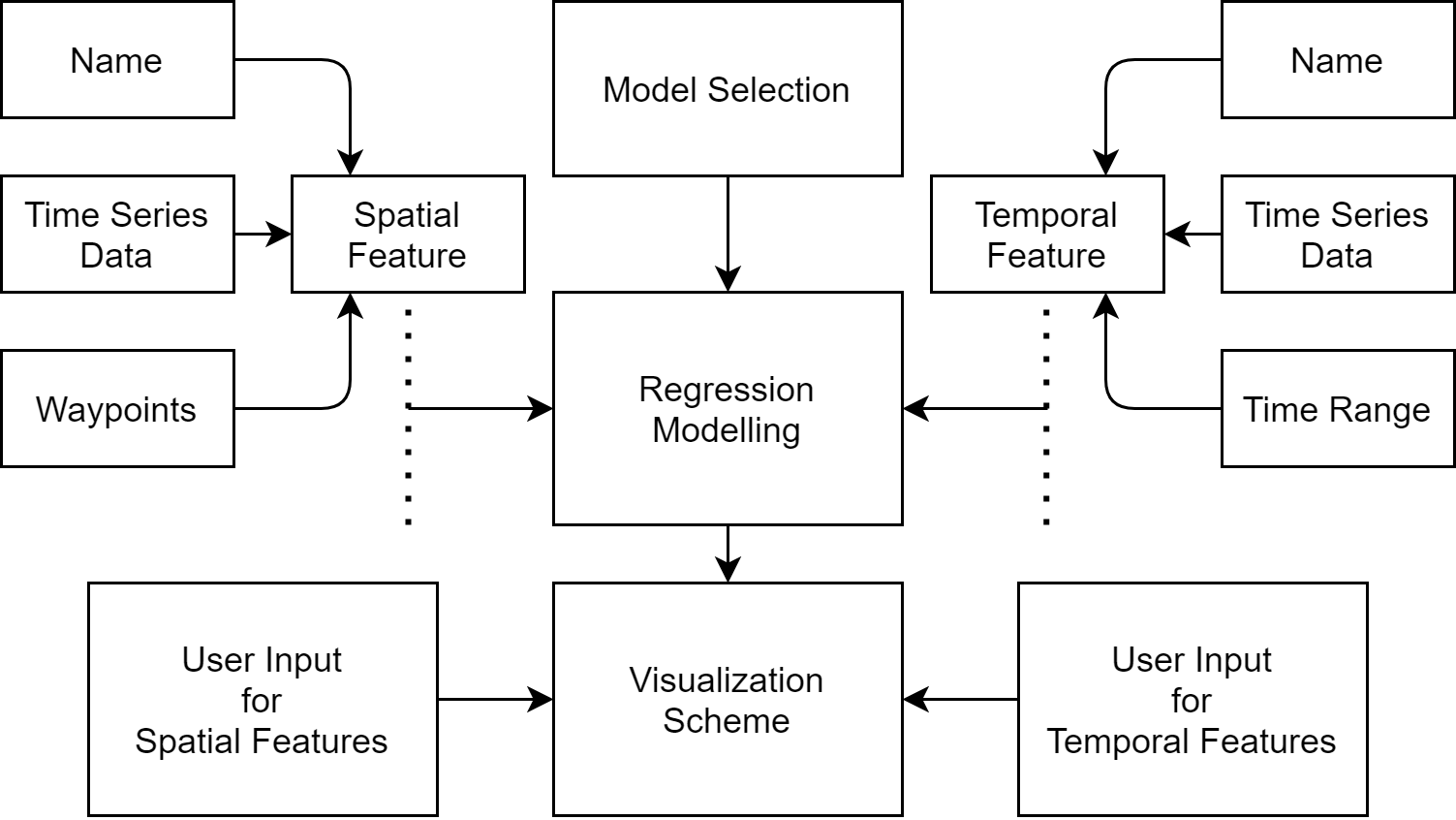}
    \caption{Processing Pipeline.}
    \label{fig:ProcessingPipeline}
\end{figure}

\par The processing pipeline of RegTraffic is shown in Figure - \ref{fig:ProcessingPipeline}. The spatial feature consists of a unique name of the feature, the corresponding time series dataset of that spatial feature, and a set of latitude and longitude as the waypoints of the route of that spatial feature. The traffic data extraction is described in \cite{Sifatul2020}. In this process, a user selects the starting and ending points of the route of interest and specifies the time range. The traffic data extraction tool gathers time-series information of the ``congestion index” of that road link every 15 minutes throughout the time range from Google Maps. The congestion index is defined by the average speed of that road link in terms of kilometers per hour. At the end of the process, the tool generates a time series dataset that has a unique name as provided by the user when adding a spatial feature in RegTraffic.

\par RegTraffic also constructs a temporal feature component with three core input values. These are the unique name of the temporal feature, the corresponding time series dataset, and the time range of that temporal feature. RegTraffic takes a set of input preferences from the user as part of the model selection. It also allows users to choose the dependent feature for the regression model. Once the regression model is built, RegTraffic passes the regression coefficients to a visualization interface where a user can input new observations for the independent features that can be both spatial and temporal.

\section{Simulation}\label{sec:Simulation}

\subsection{Spatial Feature}\label{subsec:SimulationSpatialFeature}

\par We conduct our experiment on four
connected road links in Oshawa, Ontario, Canada as shown in Figure \ref{subfig:MapJunction}. The ending point of road links 2,
3 and 4 are connected with the origin of the road link
1. A connected road network is formed by these road links. We represent the traffic congestion level of
these 4 road links as Road1, Road2, Road3, and Road4, respectively. Road1 is the dependent link where Road2, Road3 and
Road4 are the independent links that collectively affect Road1 during a specific time of the
day.

\par We collect the average traffic speed of each road link
every 15 minutes for an entire week from 12:00 am March
01, 2020, to 11:45 pm March 07, 2020. As a result, there are
a total of 672 observations over 7 days of time-series data for
each road link. Figure \ref{subfig:TimeSeries} shows the time
series of the average traffic speed of all four road links
for the first two days. The $y$ axis represents the average traffic speed in km/h, which is considered the traffic congestion
index in our analysis. We can see that the time series has a
cycle as the average traffic speed shows regular and predictable
changes that recur every day within a certain time interval. The
higher the average speed, the low the traffic congestion, and vice versa.

\subsection{Temporal Feature Extraction}\label{subsec:SimulationTemporalFeatureExtraction}

\par
\begin{figure}[t]
    \subfigure[Intersection of Simcoe and Conlin Road in Oshawa]
    {
        \includegraphics[width=0.46765\linewidth]{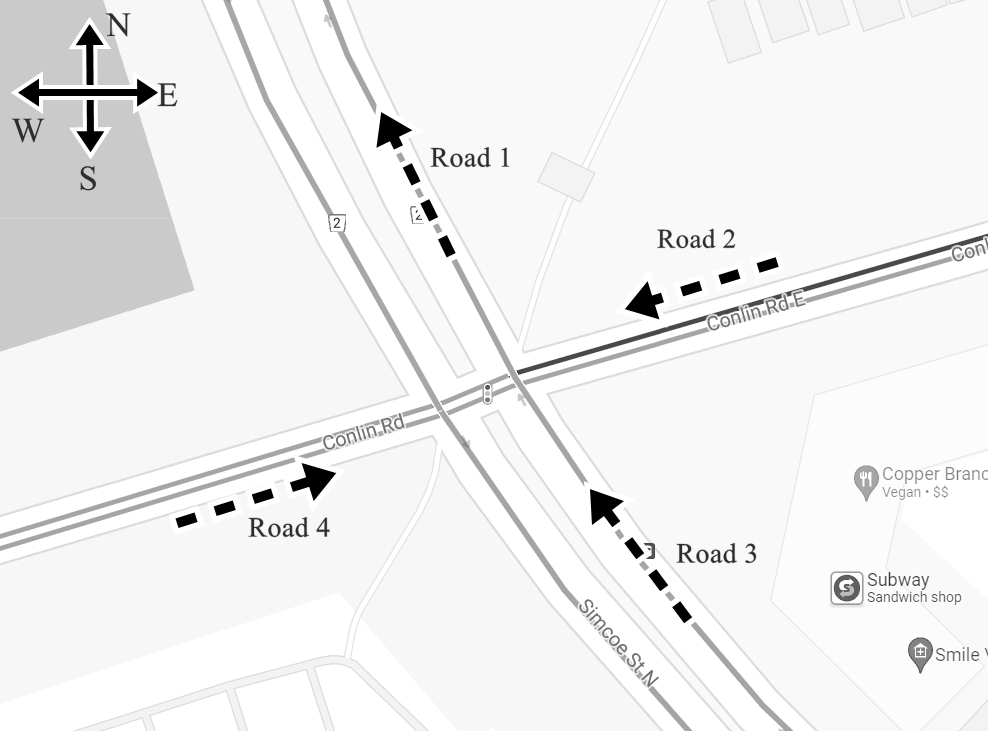}
        \label{subfig:MapJunction}
    }
    \subfigure[Time series data of 4 road links]
    {
        \centering
        \includegraphics[width=0.46765\linewidth]{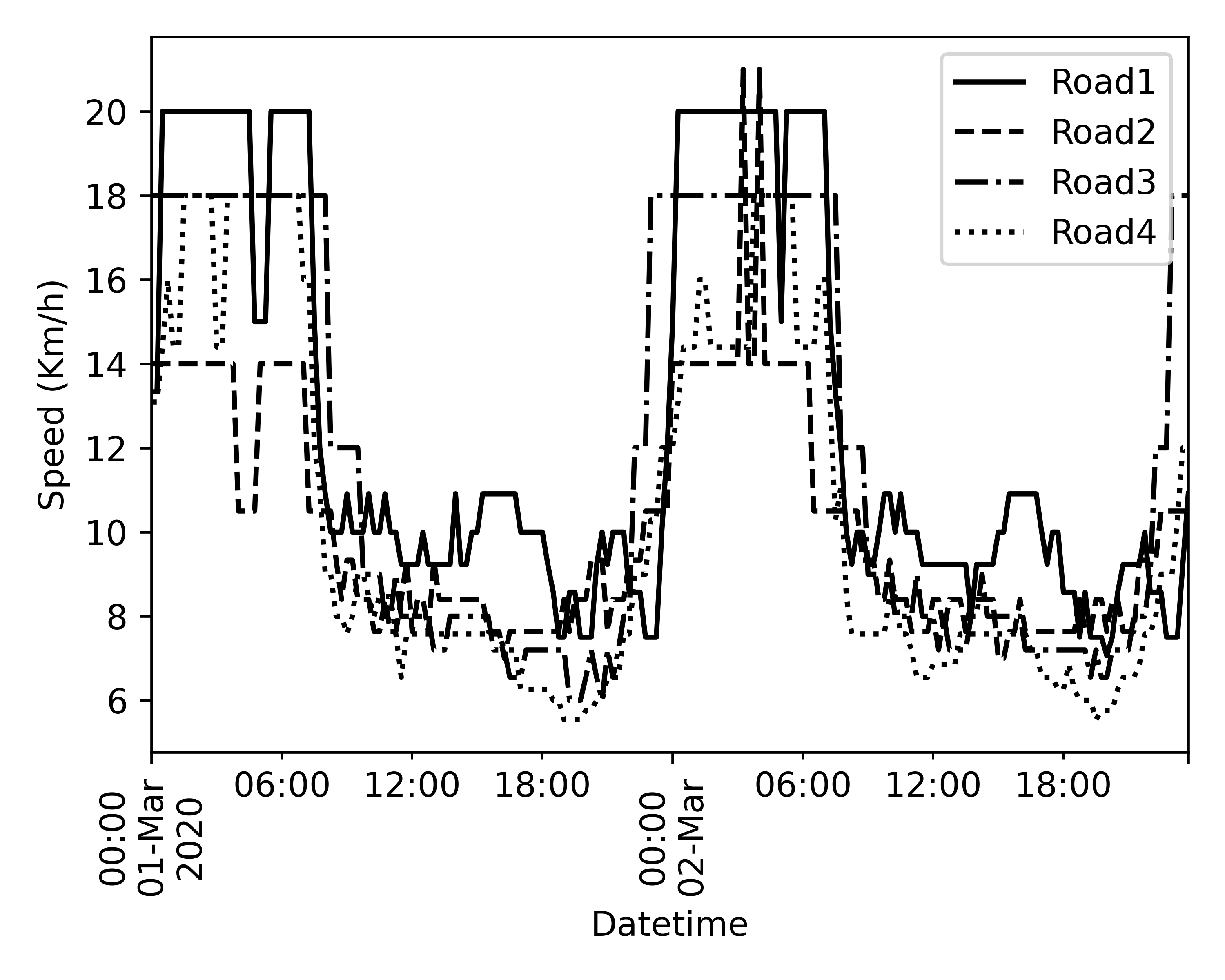}
        \label{subfig:TimeSeries}
    }
    \subfigure[Hourly average speed throughout a day]
    {
        \centering
        \includegraphics[width=0.46765\linewidth]{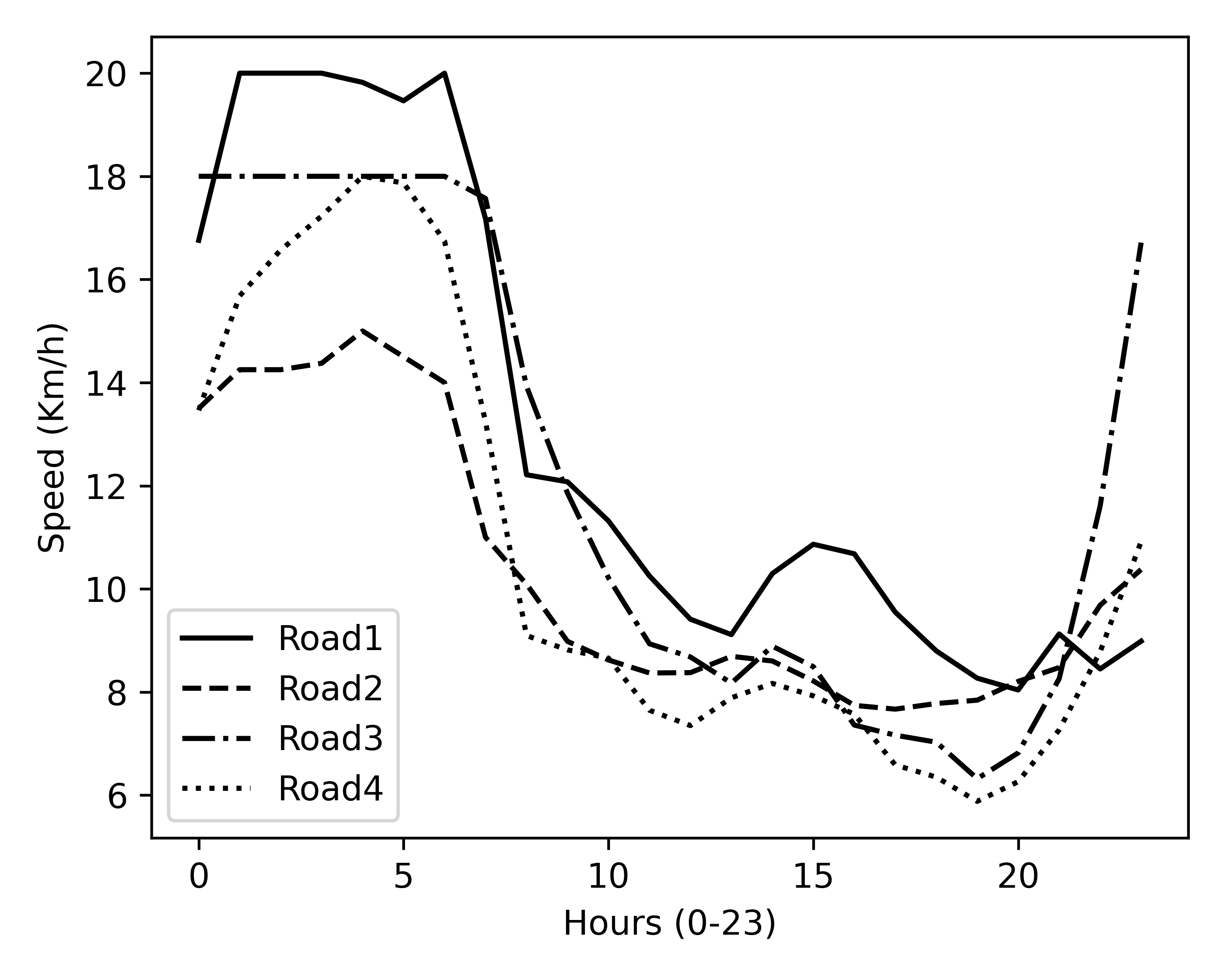}
        \label{subfig:HourlyMeanOfTrafficCongestion}
    }
    \subfigure[Identifying threshold for Peakhour]
    {
        \centering
        \includegraphics[width=0.46765\linewidth]{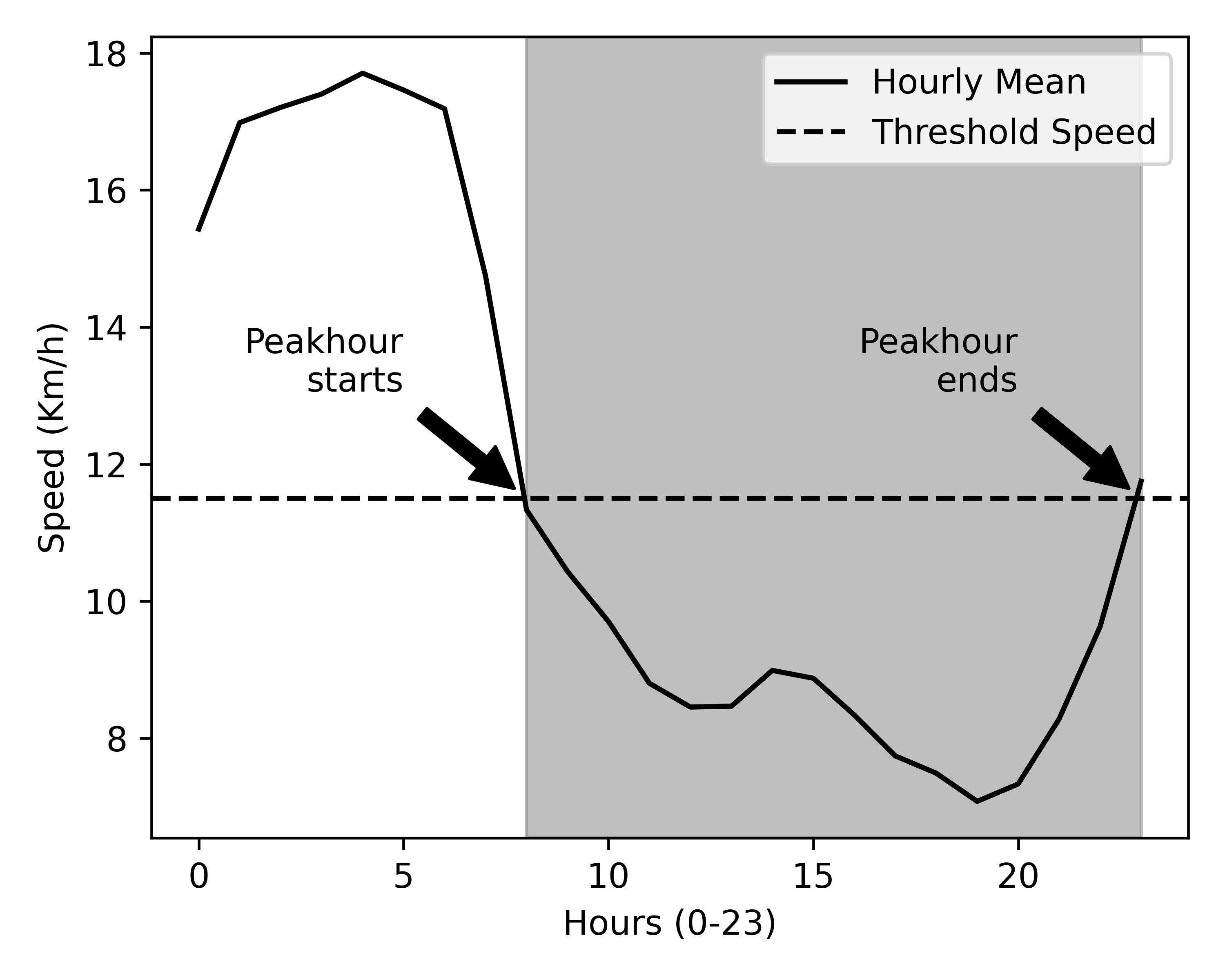}
        \label{subfig:IdentifyingThresholdForPeakhour}
    }
    \caption{Average speed throughout a day}
    \label{fig:FeatureExtractionThroughExploratoryDataAnalysis}
\end{figure}

 Figure \ref{subfig:HourlyMeanOfTrafficCongestion} shows the hourly mean of the average speed for each road link. The mean values show very little variance compared to each other as they seem to move together throughout the day. The average of the different means of all road links is plotted in Figure \ref{subfig:IdentifyingThresholdForPeakhour}. The horizontal line at a speed of 11.75 km/h divides the plot evenly and intersects with the total average speed at two points, one at daytime 8:00 and the other one at 23:00. From this exploratory data analysis, a new categorical feature called $Peakhour$ is extracted that indicates a certain time interval during a day where the average traffic speed remains below 11.75 km/h. From 9:00 am to 12:00 pm, the value of $Peakhour$ would be 1, otherwise 0. Another temporal component is considered in the analysis as a categorical variable which is $AM$. The value of $AM$ would be 1 when the meridiem is AM and 0 when it is PM.

\subsection{Regression Modeling}\label{}

\par The outcome of our Bayesian linear regression is the distribution of the model parameters. The model does not provide an exact estimate for a feature, but the mean value of the distribution can be considered as an estimate for the feature. The benefit of having a posterior probability distribution is that the model also provides an entire range of values that shows the uncertainty of the true values. The mean of a posterior probability distribution is taken as the best estimate of that model parameter. These mean estimates of these model parameters are put together to derive a new Eq. \eqref{eqn:FinalFormula}.

\begin{equation} \label{eqn:FinalFormula}
\begin{aligned}
Road1 = 7.4163*Intercept + 1.7561*AM\\
- 2.7517*Peakhour - 0.0477*Road2\\
- 0.0479*Road3 + 0.7139*Road4 + 1.7003*SD
\end{aligned}
\end{equation}

\subsection{Visualization}\label{subsec:Visualization}

\begin{figure}[t]
    \centering 
    \includegraphics[width=1.0\linewidth]{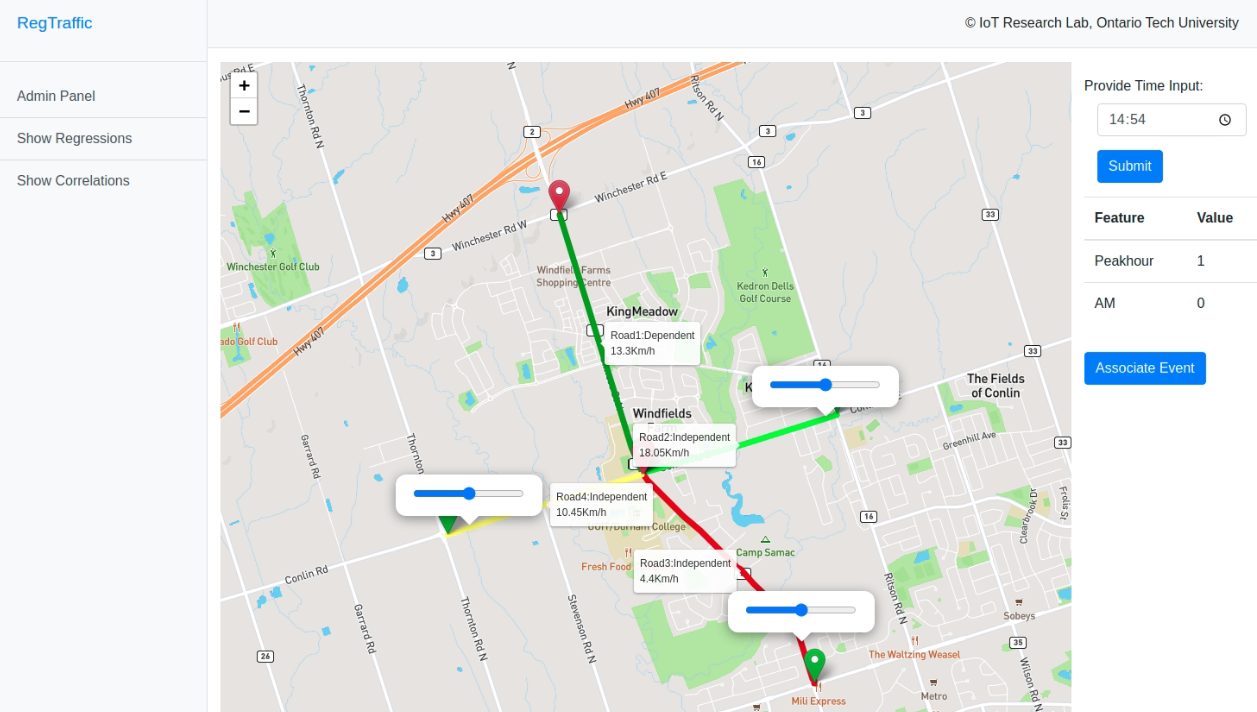}
    \caption{Regression Analysis in RegTraffic Simulator}
    \label{fig:Simulation}
\end{figure}

Figure \ref{fig:Simulation} describes a sample simulation procedure of a road intersection where $Road1$ is considered as a dependent road link and $Road2$, $Road3$ and $Road4$ are independent road links. Based on the spatial features, two new temporal features are extracted which are $Peakhour$ and $AM$. RegTraffic shows the location of the road links on an interactive geographical map where the user can provide new observations for independent road links and temporal features to predict the outcome of the dependent road link. As shown in the figure, the user sets the congestion index of  $Road2$, $Road3$, and $Road4$ to 18.05, 4.4, and 10.45 kilometers per hour, respectively. The user also needs to provide the specific time as an input for the temporal features $Peakhour$ and $AM$. RegTraffic calculates the value for the temporal features from the time input provided by the user and incorporates these values along with the input values for independent spatial features to predict the congestion index of dependent road link $Road1$. Based on the input values provided by the user, RegTraffic predicts the congestion index of the road link $Road1$, which is 13.3 kilometers per hour in this case.

\section{Performance Evaluation}\label{sec:PerformanceEvaluation}

\subsection{Test Observations}\label{subsec:TestObservation}

\par
\begin{figure}[ht]
    \subfigure[]
    {
        \includegraphics[width=0.46765\linewidth]{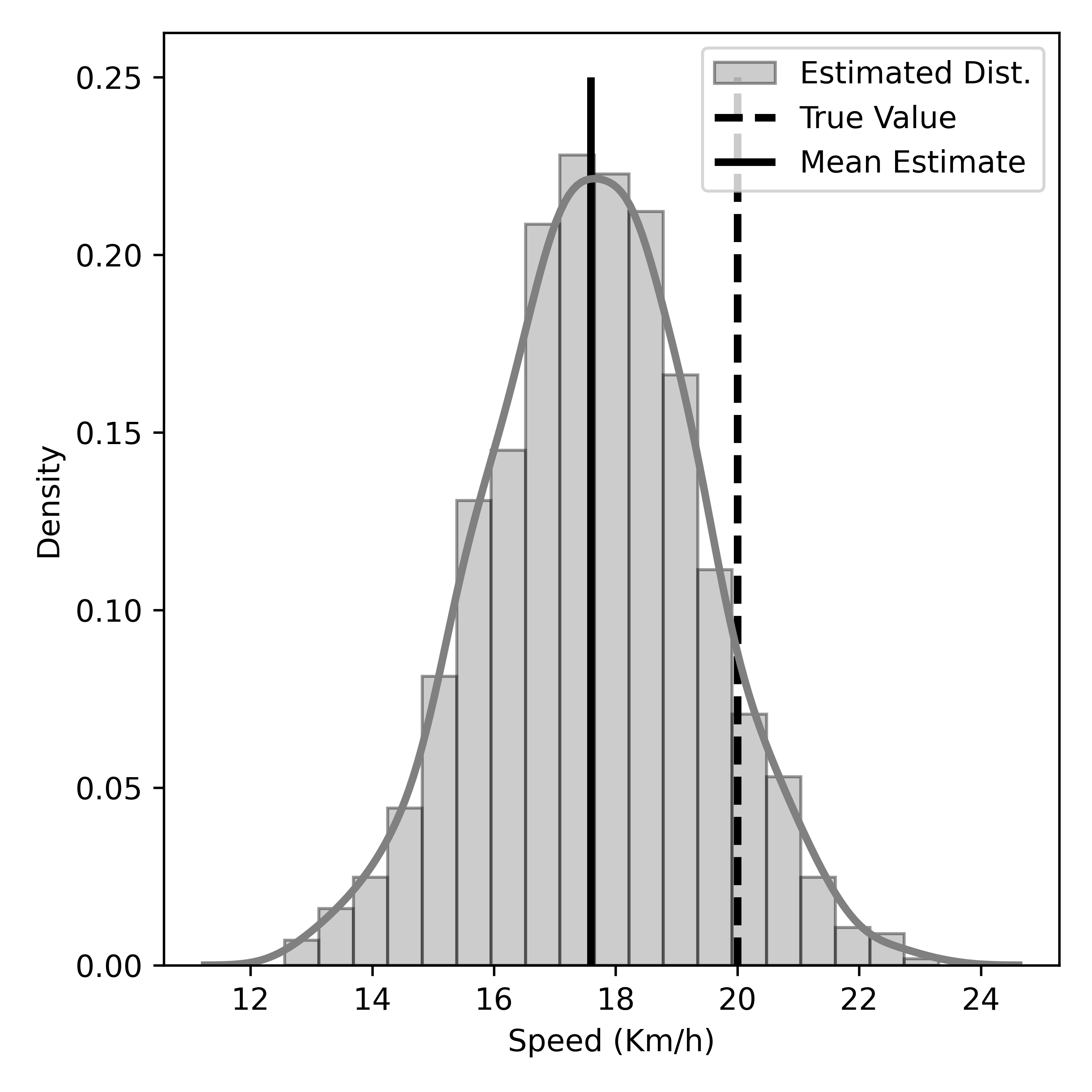}
        \label{test:first}
    }
    \subfigure[]
    {
        \includegraphics[width=0.46765\linewidth]{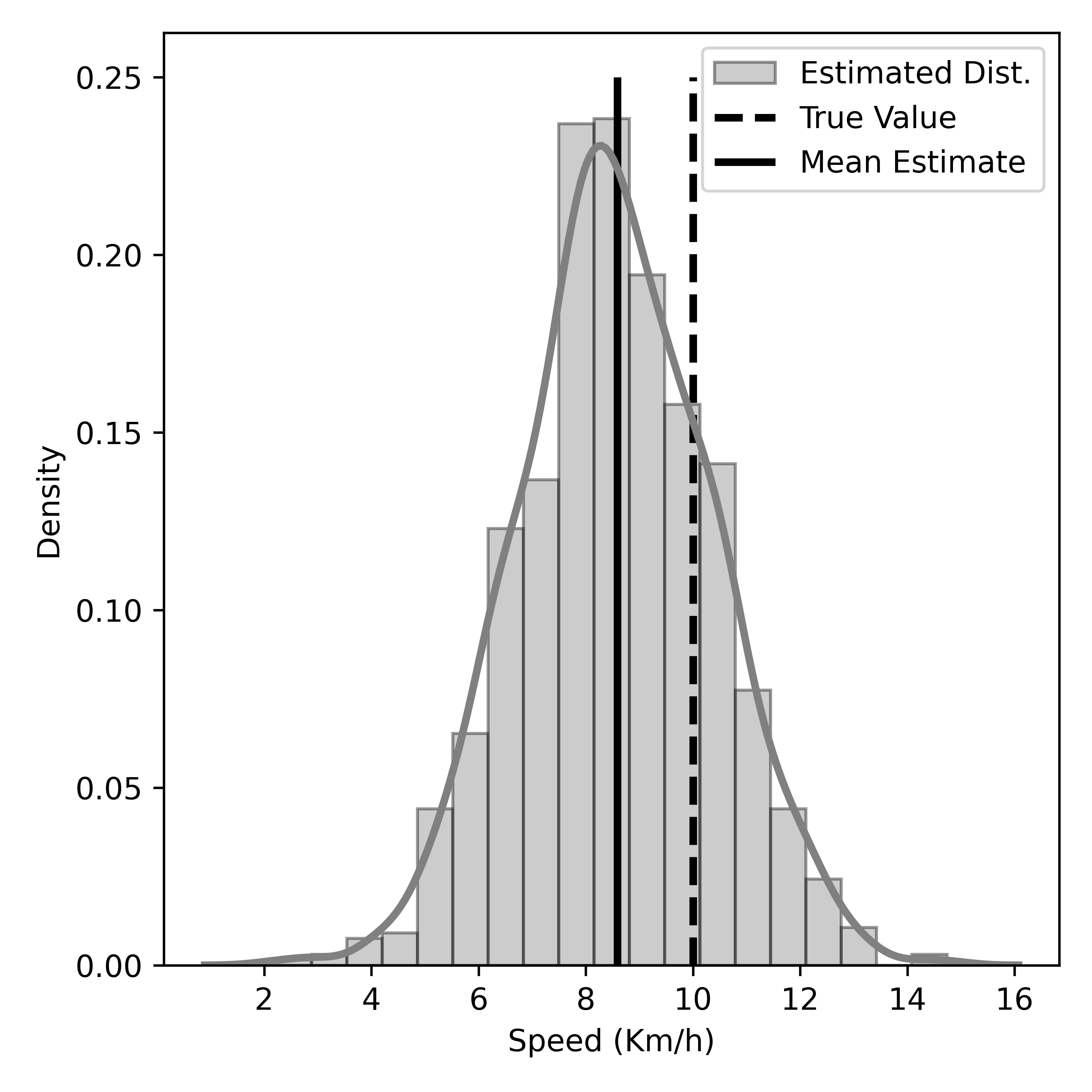}
        \label{test:second}
    }
    \subfigure[]
    {
        \includegraphics[width=0.46765\linewidth]{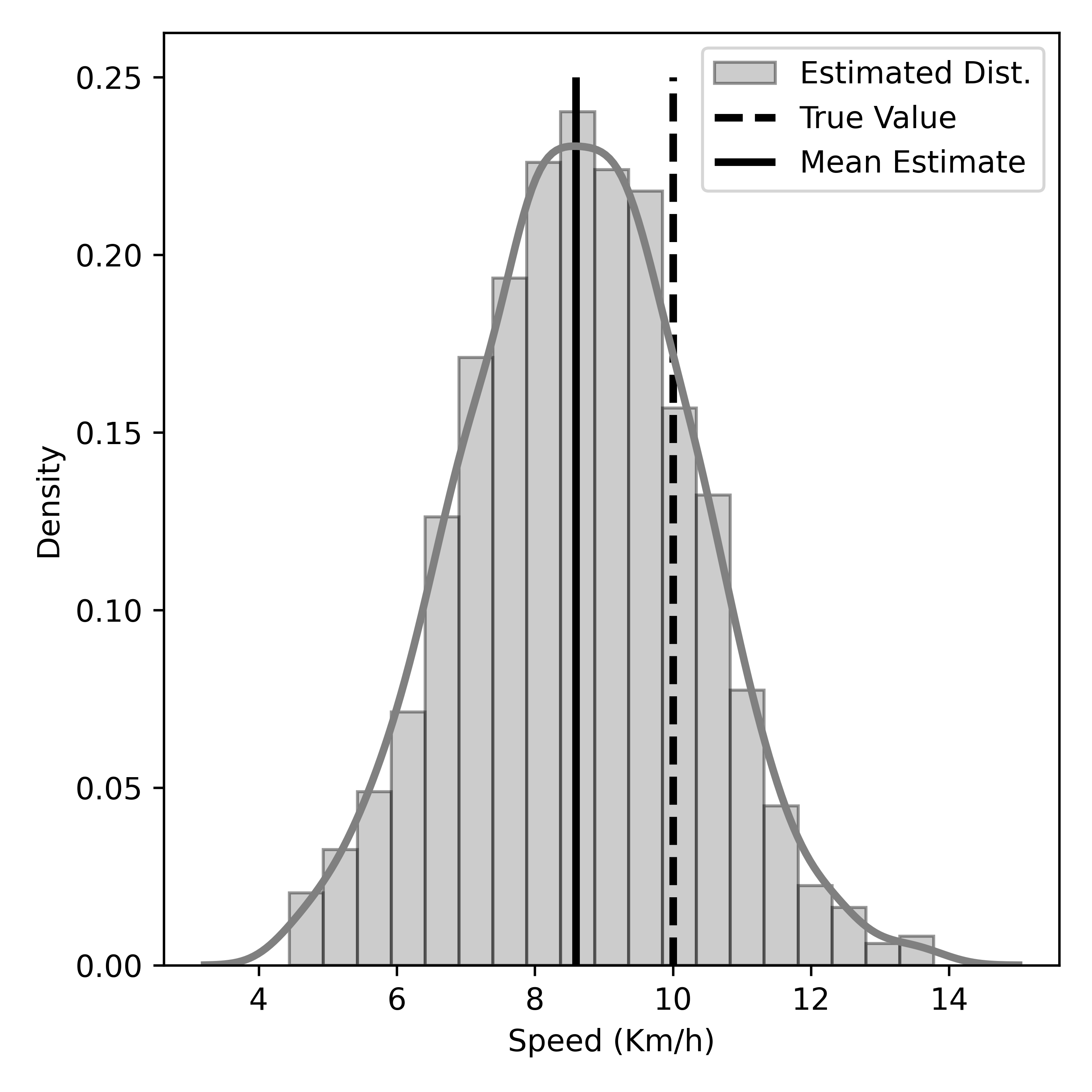}
        \label{test:third}
    }
    \subfigure[]
    {
        \includegraphics[width=0.46765\linewidth]{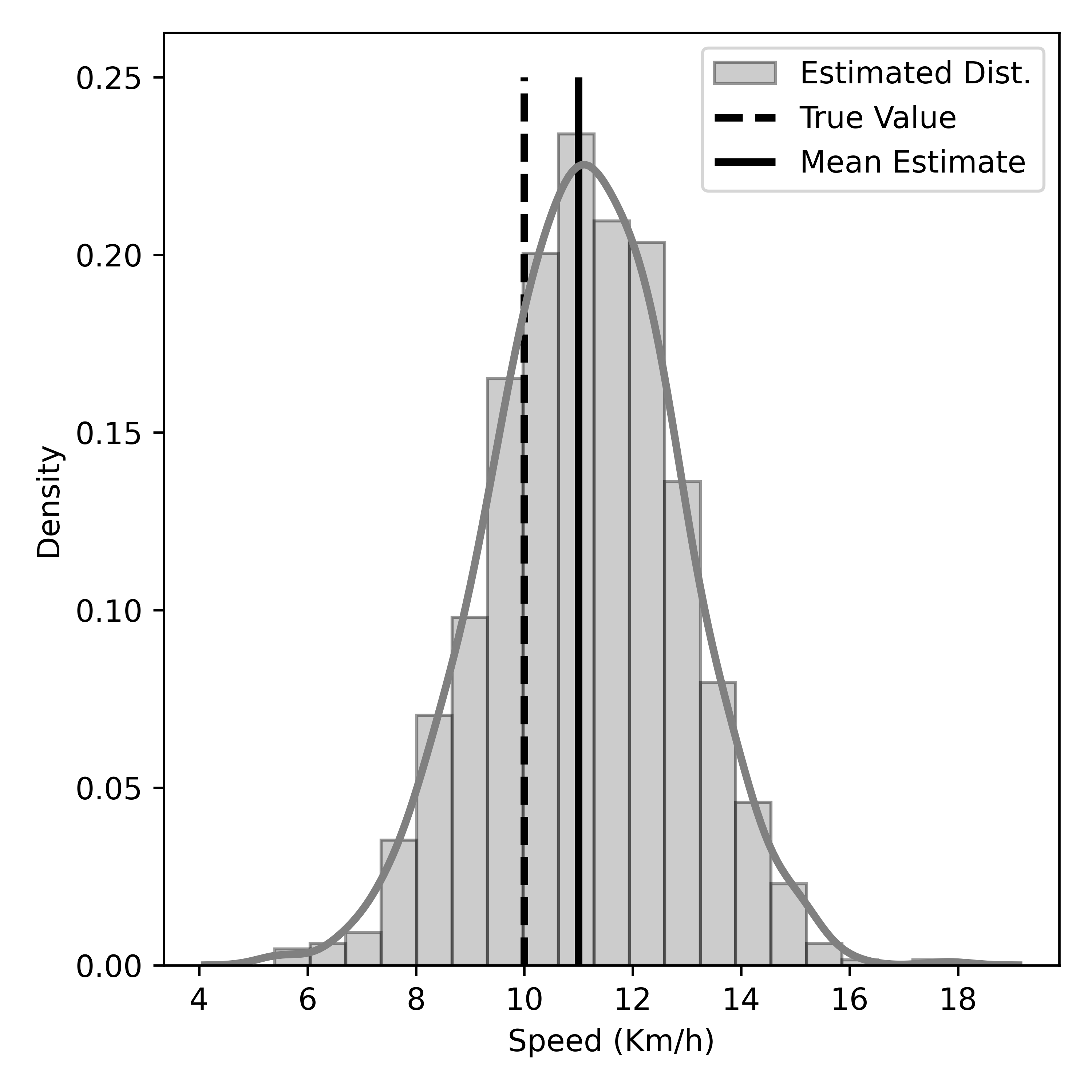}
        \label{test:fourth}
    }
    \caption{Test observations}
    \label{fig:TestObservation}
\end{figure}

 To evaluate the performance, the model is tested on a testing dataset of traffic observations. Figure \ref{fig:TestObservation} shows four random test observations from the testing dataset along with the probability density function of $Road1$. The true value of $Road1$ is represented by the dotted line and the mean of the probability distribution is represented by the straight line. The mean of the probability distribution is considered as the best estimate for the distributions. The estimated value provided by the model is very close to the true value in Figures \ref{test:first}, \ref{test:second}, \ref{test:third} and \ref{test:fourth}.

\begin{table}[t]
\centering
\caption{MODEL COMPARISON BASED ON DIFFERENT FEATURES}
\label{table:Comparison of Traffic Models}
\begin{tabular}{|c|c|c|}
\hline
 & \begin{tabular}[c]{@{}c@{}}Mean Absolute\\ Error\end{tabular} & \begin{tabular}[c]{@{}c@{}}Root Mean Squared\\ Error\end{tabular} \\ \hline
Multiple Linear Regression & 1.31269 & 1.71981 \\ \hline
Elastic Net Regression     & 1.33501 & 1.91345 \\ \hline
Bayesian Linear Regression & 1.3123  & 1.71962 \\ \hline
Baseline                   & 3.75357 & 5.09258 \\ \hline
\end{tabular}%
\end{table}

\subsection{New Observations}\label{subsec:NewObservation}

\par
\begin{figure}[t]
    \subfigure[]
    {
        \includegraphics[width=0.46765\linewidth]{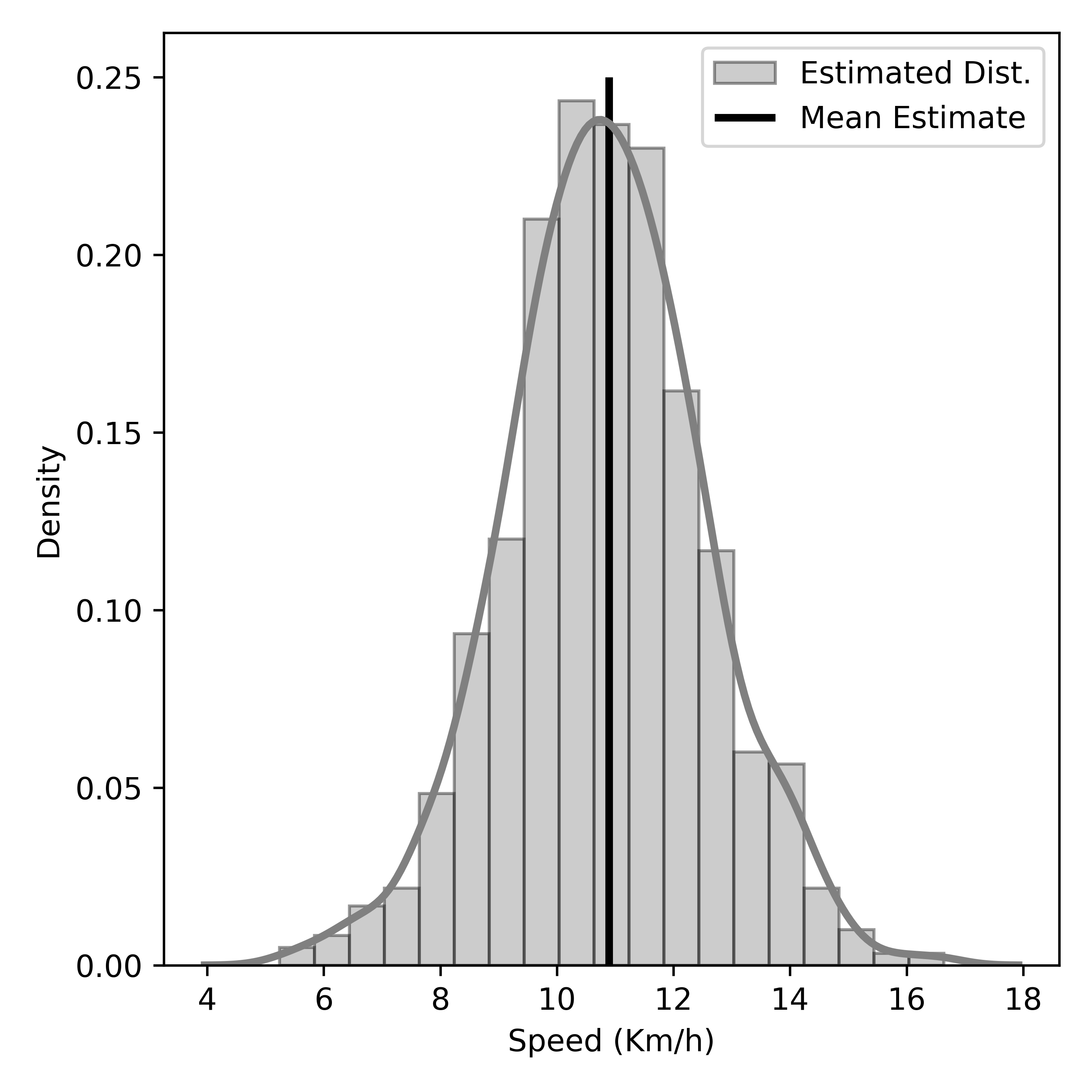}
        \label{new:first}
    }
    \subfigure[]
    {
        \includegraphics[width=0.46765\linewidth]{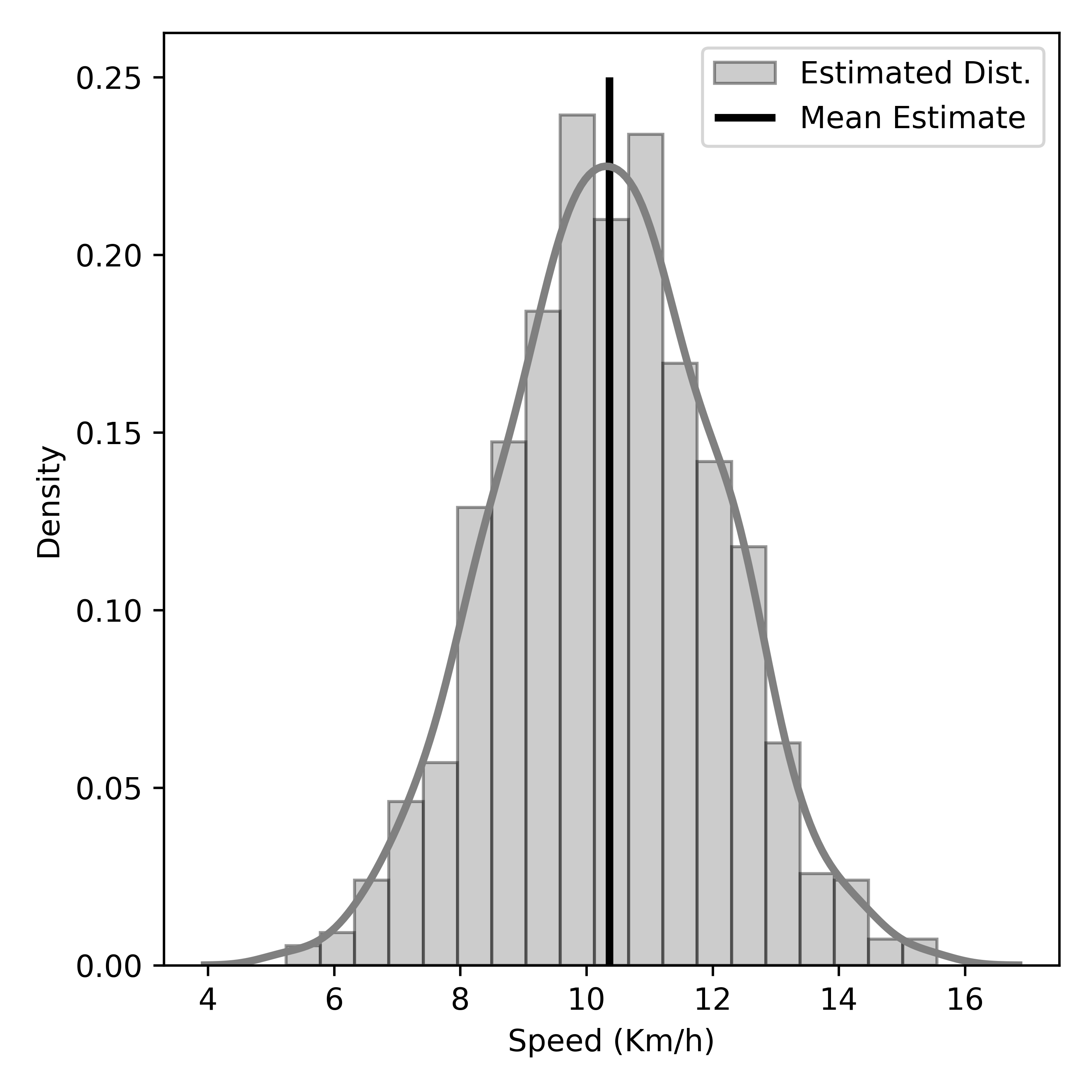}
        \label{new:second}
    }
    \subfigure[]
    {
        \includegraphics[width=0.46765\linewidth]{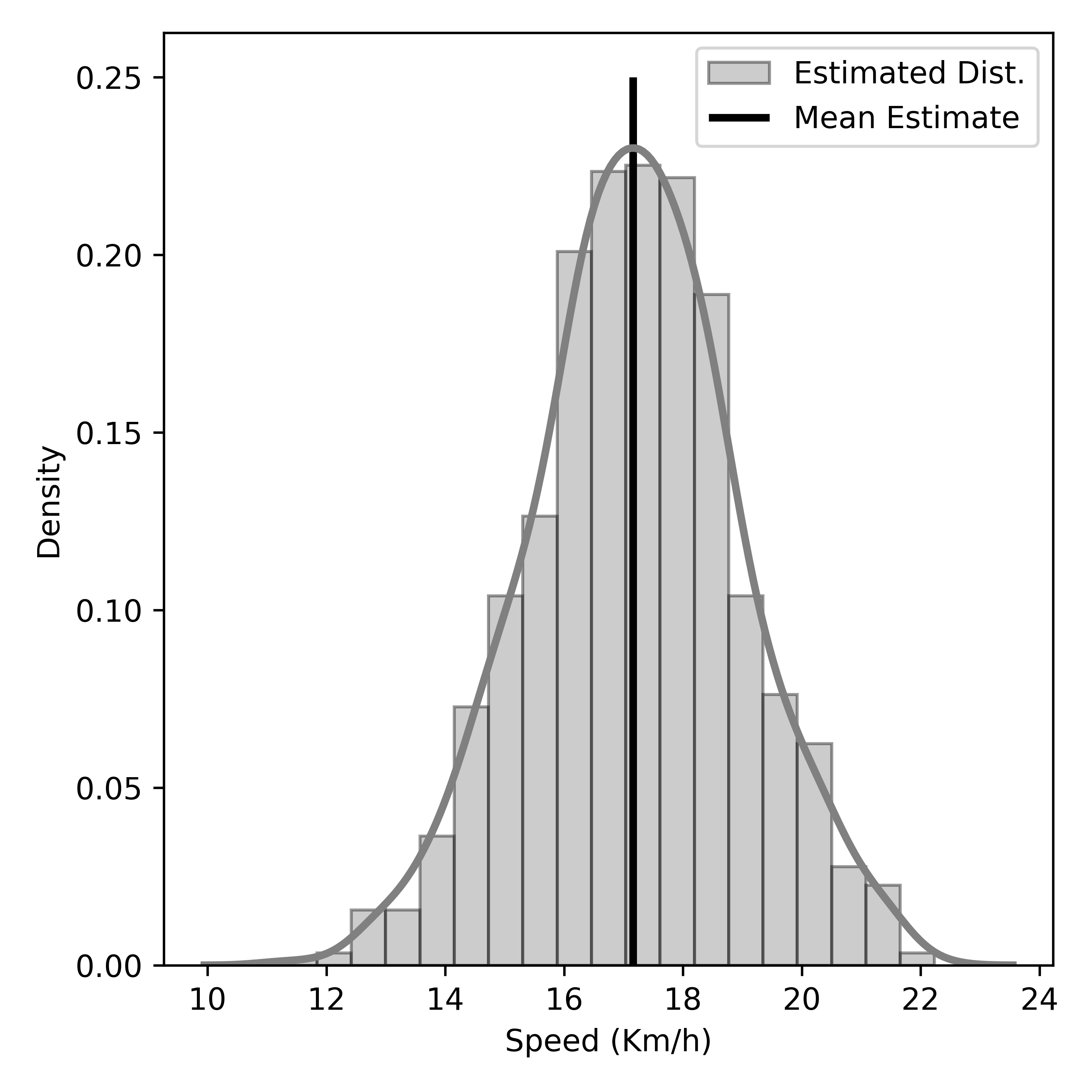}
        \label{new:third}
    }
    \subfigure[]
    {
        \includegraphics[width=0.46765\linewidth]{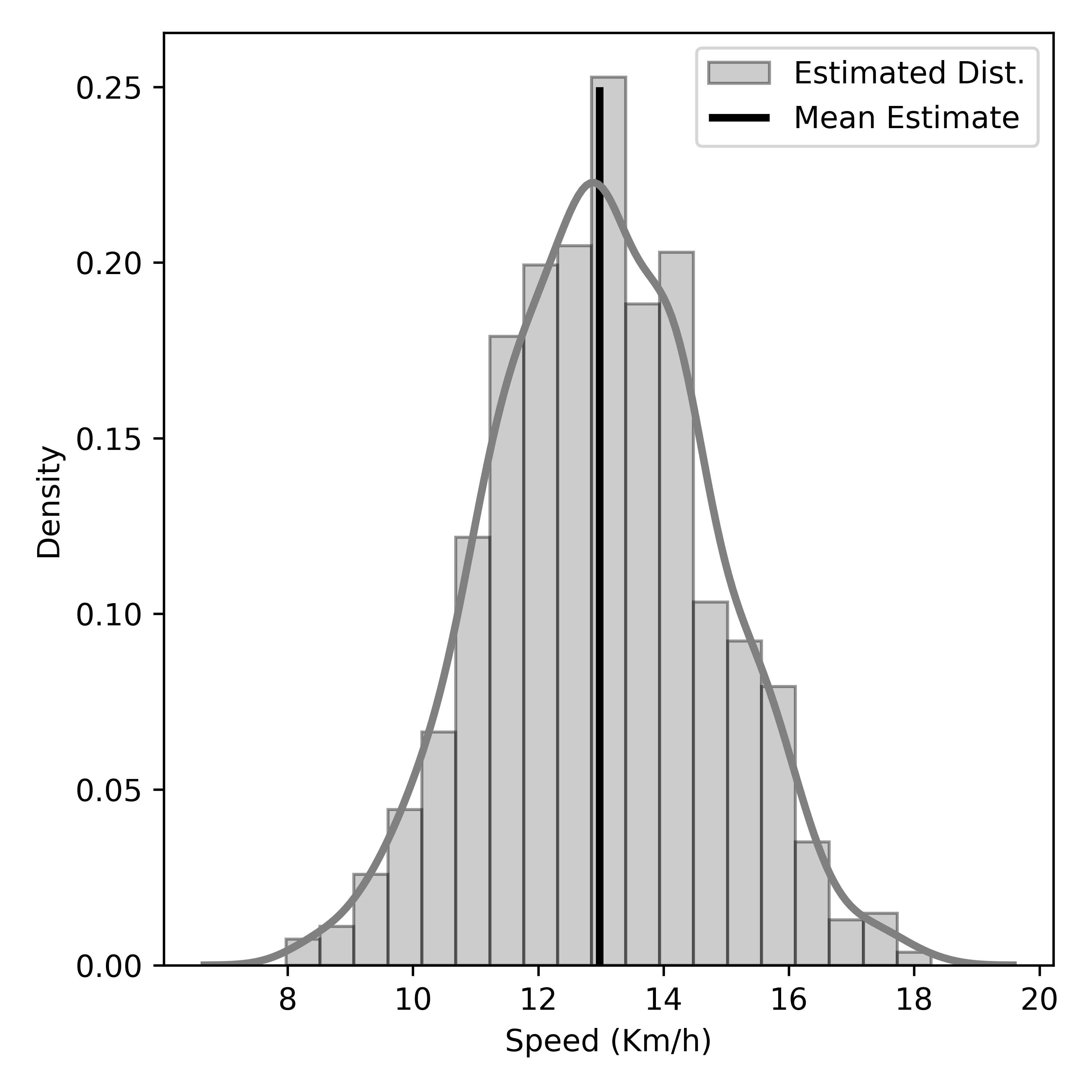}
        \label{new:fourth}
    }
    \caption{New observations}
    \label{fig:NewObservation}
\end{figure}

 To see how the model performs for new and modified observations, we test the model with a set of new observations with random values for both the spatial and temporal components as shown in Figure \ref{fig:NewObservation}. For every new observation, the model provides a new posterior distribution with the mean estimate. The vertical straight line represents the mean estimate of the posterior probability distribution for a new observation. We can see the highest probability density near the mean estimation of all posterior probability distributions as shown in Figures \ref{new:first}, \ref{new:second}, and \ref{new:third} and \ref{new:fourth}.

\subsection{Comparison With Other Approaches}\label{subsec:ComparisionWithOtherApproaches}

The performance of the Bayesian linear regression model is compared in terms of Mean Absolute Error (MAE) and Root Mean Squared Error (RMSE) with two state-of-the-art frequentist models: Multiple Linear Regression and Elastic Net Regression as shown in Table \ref{table:Comparison of Traffic Models}. We also develop a comparison baseline which is the mean of all possible observations of the traffic congestion. Here, Bayesian linear regression outperforms the state-of the-art-approaches in terms of accuracy as it has the lowest MAE and RMSE values.


\FloatBarrier

\section{Conclusion}\label{sec:Conclusion}

This paper presents RegTraffic, a new dynamic traffic simulator for spatiotemporal traffic modeling for intercorrelated road links. RegTraffic builds a regression-based spatiotemporal traffic model to predict traffic congestion of a road link depending on neighboring road links and temporal components extracted through exploratory data analysis. RegTraffic provides a dynamic interface for a user to provide new observations for independent features of the regression model and provides visualization on interactive geographical maps. The Mean Absolute Error and Root Mean Squared Error metrics are used to evaluate the performance of the regression-based predictive model integrated into RegTraffic. Performance evaluation shows that RegTraffic can effectively predict traffic congestion of intercorrelated road links. In the current version of RegTraffic, we apply a Bayesian linear regression model for better interpretation and uncertainty evaluation. In the future, we plan to enhance RegTraffic by supporting other regression-based spatiotemporal traffic modeling approaches.

\FloatBarrier


\vspace{12pt}


\end{document}